\documentclass[runningheads]{llncs}

\usepackage{eccv}

\usepackage{eccvabbrv}

\usepackage{graphicx}
\usepackage{booktabs}
\usepackage{multirow}

\usepackage[accsupp]{axessibility}  % Improves PDF readability for those with disabilities.

\usepackage{hyperref}

\usepackage{orcidlink}

\begin{document}

\title{Identity-Consistent Diffusion Network for Grading Knee Osteoarthritis Progression in Radiographic Imaging} 

\titlerunning{IC-RDN for Grading Knee Osteoarthritis Progression}

\author{Wenhua Wu\inst{1}\orcidlink{0000-0001-7488-5487} \and
Kun Hu\inst{1,}\thanks{Corresponding author.} \orcidlink{0000-0002-6891-8059} \and
% {Kun Hu \inst{1}}^{(\textsuperscript \Letter)} \orcidlink{0000-0002-6891-8059} \and
Wenxi Yue \inst{1}\orcidlink{0000-0002-8270-417X} \and
Wei Li \inst{1}\orcidlink{0000-0003-4731-3226} \and
Milena Simic \inst{1} \and
Changyang Li \inst{2}\orcidlink{0000-0001-9400-3998} \and
Wei Xiang \inst{3}\orcidlink{0000-0001-9400-3998} \and
Zhiyong Wang \inst{1}\orcidlink{0000-0002-8043-0312} 
}

\authorrunning{W. Wu et al.}

\institute{The University of Sydney, Darlington NSW 2008, Australia \\
\email{\{wenhua.wu, kun.hu, wenxi.yue, weiwilson.li, milena.simic, zhiyong.wang\}@sydney.edu.au} \and
Sydney Polytechnic Institute Pty Ltd, Sydney, NSW, Australia\\
\email{chris@ruddergroup.com.au} \and
School of Computing, Engineering \& Mathematical Sciences, La Trobe University\\
\email{w.xiang@latrobe.edu.au}}
\maketitle

\begin{abstract}
Knee osteoarthritis (KOA), a common form of arthritis that causes physical disability, has become increasingly prevalent in society. Employing computer-aided techniques to automatically assess the severity and progression of KOA can greatly benefit KOA treatment and disease management. Particularly, the advancement of X-ray technology in KOA demonstrates its potential for this purpose. Yet, existing X-ray prognosis research generally yields a singular progression severity grade, overlooking the potential visual changes for understanding and explaining the progression outcome. 
Therefore, in this study, a novel generative model is proposed, namely Identity-Consistent Radiographic Diffusion Network (IC-RDN), for multifaceted KOA prognosis encompassing a predicted future knee X-ray scan conditioned on the baseline scan. 
Specifically, an identity prior module for the diffusion and a downstream generation-guided progression prediction module are introduced. 
Compared to conventional image-to-image generative models, identity priors regularize and guide the diffusion to focus more on the clinical nuances of the prognosis based on a contrastive learning strategy.  
The progression prediction module utilizes both forecasted and baseline knee scans, and a more comprehensive formulation of KOA severity progression grading is expected.  
Extensive experiments on a widely used public dataset, OAI, demonstrate the effectiveness of the proposed method. 
% \footnote{Our code will be publicly available. }
\end{abstract}

\section{Introduction}
\label{sec:intro}

\begin{figure*}[!hbt]
\centering
\includegraphics[width=0.95\textwidth]{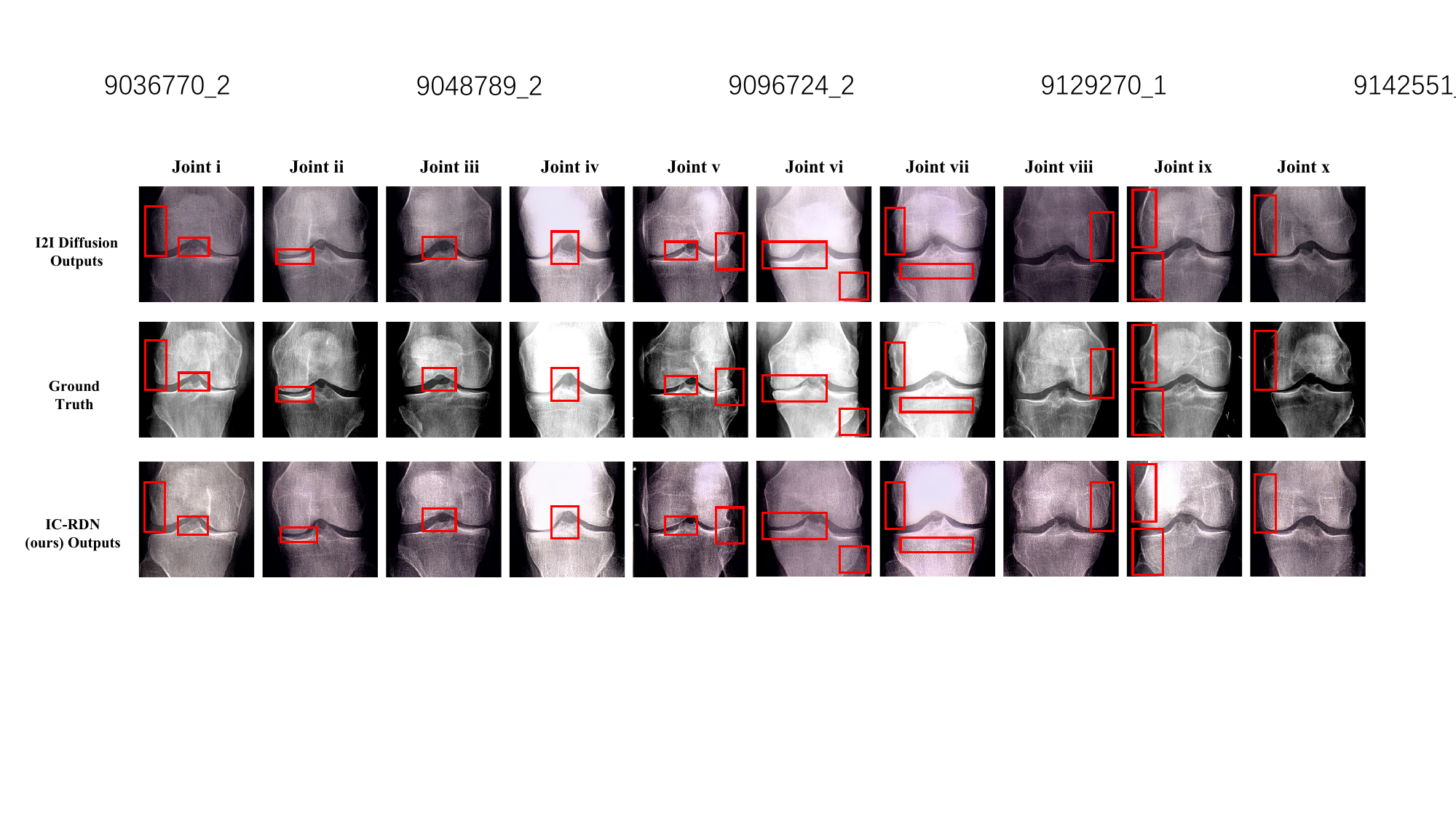}
\caption{Pipeline for KOA severity progression grading, encompassing a) predicted future knee joint X-ray scan, and b) predicted KOA progression severity grade. }
\label{fig:intro_fig}
\end{figure*}

Knee osteoarthritis (KOA) is a common form of arthritis, mainly affecting knee joints with the symptoms of knee joint swelling, tenderness, and bone spurs. As the major factor of physical disability, KOA has recently shown an increasing prevalence in society, especially among the elders \cite{cui2020global}. Currently, there are rare treatments for curing KOA without physical surgery (total knee replacement) \cite{guermazi2009plain}. Therefore, to relieve the pain and control the progression of KOA, the diagnosis and prognosis of the disease are significant, using different clinical assessment methods, such as medical imaging scans and patients' clinic questionnaires. In specific, the imaging scans are utilized to detect the structural changes within the knee joint areas, which indicates the characteristics of KOA. Typically, the severity of KOA is measured by the Kellgren and Lawrence (KL) grading scale system, which encompasses five different levels, from grade 0 (healthy cases) to grade 4 (highly severe cases) \cite{kellgren1957radiological}. 

Compared to the extensive time and human resources required for specialists to detect changes, employing computer-aided techniques to automatically assess the severity and progression of KOA can significantly reduce time costs. This aids in the prompt diagnosis and prognosis of KOA at early stages \cite{brahim2019decision, hu2022adversarial}. 
Methods for grading the severity and progression of KOA have evolved from traditional machine learning approaches \cite{shamir2008knee, shamir2009early}  to advanced deep learning techniques \cite{chen2019fully, tiulpin2020automatic}.  These modern approaches utilize various forms of input, including single images, multiple images, and images combined with demographic information (such as age, gender, BMI, and the WOMAC questionnaire) \cite{von2020towards, nasser2022discriminative, zhang2020attention, tiulpin2019multimodal}.
Particularly, the advancement of X-ray technology and its application in KOA significantly improves the capability to detect and monitor the progression  \cite{guan2022deep, hirvasniemi2023knee, hu2022adversarial}. Meanwhile, utilising the generative AI methods in the KOA analysis field focuses on augmenting the knee radiological dataset to improve the grading process \cite{prezja2023exploring}. 
Yet, existing X-ray prognosis research generally yields a singular progression severity grade using the baseline X-ray scan \cite{hirvasniemi2023knee, guan2022deep, almhdie2022prediction}, overlooking the potential visual changes for understanding and explaining the progression outcome. However, the visually predicted outcome is also crucial in clinical practice to explain the disease trajectory. 

In addition to enhancing these methods, we hope the longitude visual content will also aid clinicians in their decision-making by providing further explainability and as domain knowledge to guide the algorithms. For the methodology part, previous findings revealed that GANs yield poor bone structures and inconsistency in patients' identity information, which motivated us to devise a diffusion-based method with identity information. Identity patterns are crucial with a principle widely utilized in the forensic field. We anticipate that incorporating such information improves the consistency between baselines and synthesized longitude images.

Therefore, in this study, a novel deep generative model, namely Identity-Consistent Radiographic Diffusion Network (IC-RDN), is proposed for multifaceted KOA prognosis. IC-RDN creates a predicted future knee X-ray scan conditioned on the baseline scan and a future KOA severity grade as illustrated in Fig. \ref{fig:intro_fig}. 
Specifically, an identity prior module for guiding the diffusion and a downstream generation-guided progression prediction module are introduced. 
Compared to a conventional image-to-image generative model,  identity priors regularize and guide the diffusion to focus more on the clinical nuances related to the prognosis based on a contrastive learning strategy.  
Using both forecasted and baseline knee scans, we anticipate a more comprehensive and enhanced formulation of KOA severity progression measured in grades with the progression prediction module.  
Extensive experiments on a widely used public dataset, OAI \cite{lester2008clinical}, demonstrate the effectiveness of the proposed method. 
%The future scans obtained can intuitively contribute to an improved prognosis and be further leveraged for enhanced diagnostic outcomes.

The key contributions of this study can be summarised in three-fold:

\begin{itemize}
 \item A novel deep generative model - IC-RDN is proposed for multifaceted KOA prognosis prediction for the first time, encompassing radiographic visual predictions and grading forecasting. 
 \item An identity prior module for persisting patient's identity information on bone X-ray images is proposed to guide the diffusion process focusing on clinical nuances. 
 \item Comprehensive experiments with the public OAI dataset demonstrate the effectiveness of the proposed method. 
\end{itemize}

\section{Related Works}

\subsection{Computer-Aid KOA Severity Grading}
\label{subsec:liter_severity_grading}

Existing studies in grading the severity of KOA can be categorised into KOA detection and KOA multi-class grading \cite{nasser2022discriminative, wu2023self}. The KOA detection focuses on categorising several or combined severity grades from the whole grading system to achieve a coarse-grained classification. 
%To this end, deep learning methods among CNNs, GNN, and transformer-based algorithms show promising results in grading the severity of KOA. Specifically, 
For instance, a graph neural network has been designed to utilize the shape boundary information of knee bones for KOA detection \cite{von2020towards}. 
In addition to shape, texture features have been independently utilized for KOA detection to examine the internal pixel relationship \cite{ribas2022complex}. 
To leverage both textural and shape information, the features extracted from the shallow and deeper layers of a DenseNet model are combined, where the shallow layers contain more shape patterns and the deeper layers focus on textural information \cite{nasser2022discriminative}.

Multiclass grading, on the other hand, concentrates more on exploring each grading severity level in the system. 
As the KL grades can be considered a continuous measurement, with convolution neural networks, both regression and classification strategies are studied \cite{antony2016quantifying, antony2017automatic}. 
Likewise, an ordinal loss function is proposed to formulate the continuous nature of KL grades \cite{chen2019fully}. 
In parallel, the attention mechanisms identify the significant areas related to KOA multiclass grading~\cite{gorriz2019assessing, zhang2020attention}. 
In \cite{jain2024knee}, attention mechanisms are studied to combine and fuse multi-scale features from a network. 
Furthermore, to take advantage of the symmetrical structure of knee joints, a Siamese Network is employed to investigate the relationship between both knee joints. This approach facilitates the analysis of intricate KOA patterns in both fully supervised and semi-supervised manners, addressing the challenge posed by the scarcity of annotated datasets \cite{tiulpin2018automatic, nguyen2020semixup}. Likewise, a self-supervised method has been introduced. This method leverages both X-ray and MRI images to expand the pool of positive pairs in contrastive learning settings, offering a robust solution to dataset limitations \cite{wu2023self}. 

\subsection{Computer-Aid KOA Longitude Progression Grading} 
\label{subsec:liter_progress_grading}

Apart from KOA severity grading, predicting the progression of the disease is also significant for medical treatment. Existing works mainly focus on different aspects in predicting the progression of KOA: 1) Knee joint pain status; 2) KOA risk factor prediction; 3) Joint space narrowing (JSN) and cartilage structural change prediction; 4) KL grade prediction; and 5) total knee replacement surgery (TKR) prediction. %And the classes for measuring the KOA progression are defined individually among these studies. 

Early studies on grading the progression of KOA mainly rely on conventional machine learning algorithms. In specific, the feature selection is leveraged in filtering the significant features contributing to the progression grades \cite{ntakolia2021prediction}. Parallelly, the comparison among different machine learning classifiers is conducted to select the algorithm with better progression grading performance \cite{widera2020multi}. \cite{cheung2021superiority} utilises the XGBoost model to predict the knee joint width as a KOA progression factor. To take advantage of various classification algorithms, the ensemble strategy is deployed among RandomForest, logistic regression, and support vector machine \cite{jamshidi2021machine}. 
These machine learning-based methods show some encouraging results in feature engineering, where the extracted features are combined and utilized with methods such as multilayer perceptrons (MLPs) to predict the cartilage volume from MRI scans \cite{hafezi2017prediction}. 

With the powered deep learning methods explored in both general and medical fields, end-to-end approaches have shown promising results in KOA progression grading. An MLP is deployed as a classifier to predict progression grades \cite{chan2021machine}. Neural Networks, otherwise, are utilised to predict pain status via extracting information from radiographics and demographics \cite{guan2022deep, pierson2021algorithmic}. Likewise, CNN-based methods are also leveraged in analysing various KOA risk factors \cite{chan2021machine} and predicting the JSN progression \cite{almhdie2022prediction, halilaj2018modeling}. To explore the KOA evolutionary trajectory, an Adversarial Evolving Neural Network is proposed to involve the estimated KOA patterns at different grades as auxiliary information~\cite{hu2022adversarial}. 
Although predicting the direct KOA symptoms shows a clearer progression measurement, grading the progression KL grades and predicting TKR are also able to illustrate the progression in the clinical assessment domain. A binary classification model is designed based on whether the patient needs to take TKR as a future event \cite{tolpadi2020deep, jamshidi2021machine, hirvasniemi2023knee}. 

Despite using medical imaging only, the information from multiple modalities, including imaging scans, clinical questionnaires, and demographics, are leveraged in a multimodal method to analyse the KOA progression comprehensively \cite{tiulpin2019multimodal}. However, these studies only produce a direct outcome for progression grading ignoring the visual information from the given medical imaging techniques. Therefore, including visual information as an additional modality for analysing comprehensive multifaceted progression patterns is attractive to be explored.

\begin{figure*}[!htb]
\centering
\includegraphics[width=0.99\textwidth]{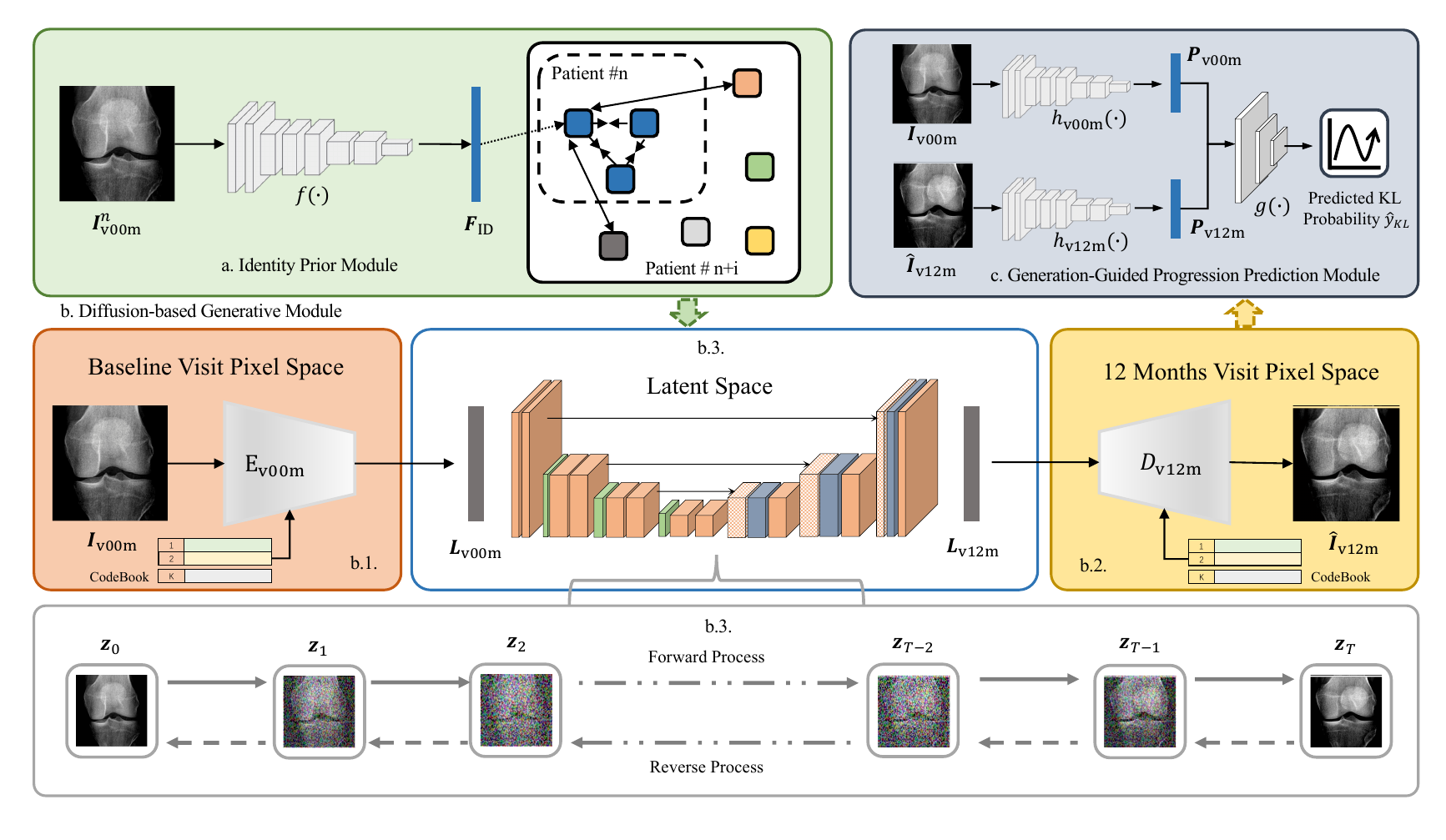}
\caption{Illustration of the proposed IC-RDN method. (a) An identity prior module formulates identity representation, which guides the diffusion process for the generation. (b) An I2I image generative network to forecast a future 12-month X-ray scan. (c) A downstream KOA progression prediction network, taking the baseline and the predicted 12-month X-ray scans as inputs. 
}
\label{fig:diffusion_model}
\end{figure*}

\section{Methodology} \label{sec:gen_method}

As shown in Fig. \ref{fig:diffusion_model}, the proposed IC-RDN consists of three modules: 1) an identity prior module extracts patient's identity representations; 2) a diffusion-based generative module produces the future X-ray images based on the current-stage X-ray scans and the associated identity representations; and 3) a down-stream generation-guided progression prediction module that predicts the progression severity using both current and generated future X-ray images to provide comprehensive multifaceted progression patterns. 

\subsection{KOA Progression Radiographic Generation}

This study incorporates X-ray medical imaging scans to predict the progression of KOA. The objective is to forecast the severity of KOA progression from two perspectives, encompassing severity grading score and visual information (i.e., a generated knee X-ray scan in the future - 12-month in this study).  
Initially, radiographic images for joints are captured during the baseline (first) X-ray scan, referred to as the baseline visit or 0-month visit. Patients are then asked to undergo a follow-up X-ray scan after 12 months for longitudinal tracking.  
Specifically, the $i^\text{th}$ longitude X-ray joint pairs from the dataset is denoted as $\mathbf{I}_{\text{visit}}^i \in \{\mathbf{I}_{\text{baseline}}^i, \mathbf{I}_{\text{12-month}}^i \}$, where $\mathbf{I}_{\text{visit}}^i \in \mathbb{R}^{C\times W\times H}$ and $C$, $W$ and $H$ are the number of channels, the width and the height of the X-ray joint images. For notation simplicity,  $\mathbf{I}_{\text{Baseline}}^i \text{ and } \mathbf{I}_{\text{12 months}}^i$ can be shortened as $\mathbf{I}_{\text{v00m}}^i \text{ and } \mathbf{I}_{\text{v12m}}^i$. And the generated 12-month X-ray image for the $i^{th}$ joint pair is denoted as ${\hat{\mathbf{I}}_{\text{12 months}}^i}$ or ${\hat{\mathbf{I}}_{\text{v12m}}^i}$. The corresponding severity grade for the 12-month visit is indicated by $y_{KL}^i$ and the grades produced by the prediction model are shown by $\hat{y}_{KL}^i$. 

% change the sign of generated image --> head

\subsection{Diffusion-based Visual Prognosis}

The diffusion-based generative module in the IC-RDN is designed to produce 12-month X-ray images from baseline visit images for understanding and explaining the potential visual changes in terms of KOA patterns. 
This incorporates a T-step Denoising Diffusion Probabilistic Model (DDPM) \cite{ho2020denoising} under an image-to-image transition manner. 
The DDPM consists of two primary processes: the forward (or diffusion) process and the reverse process. 
Typically, in the forward process, the original data $x_0 \sim q_{data}(x_0)$ is transformed towards an isotropic Gaussian distribution. 
However, within our image-to-image diffusion model, the aim is to map the data $x_0 \sim q_{data}(x_0)$ to $x_T$, where, in our context, $x_0$ represents a 12-month visit X-ray image $\mathbf{I}_{\text{v12m}}^i$, and  $x_T$ corresponds to the baseline visit X-ray images $\mathbf{I}_{\text{v00m}}^i$. 
This transformation is modelled as a fixed Markov Chain of length $T$. 
Subsequently, the reverse process mirrors the forward process, aiming to predict  
$x_0$ from $x_T \sim q_{data}(x_T)$. 

However, executing both processes directly in pixel space is highly resource-intensive, necessitating substantial memory and computational power. To mitigate this, a VQ-GAN-based encoder (see Fig. \ref{fig:diffusion_model} (b.1.)) encodes $x_0$ into a latent space $z_0$, wherein both the forward and reverse processes (see Fig. \ref{fig:diffusion_model} (b.3.)) are performed to generate $z_T$. A VQ-GAN-based decoder (depicted in Fig. \ref{fig:diffusion_model} (b.2.)) then decodes  $z_T$ back from the latent space to the pixel space ($x_T$). The objective of this stable diffusion-based approach \cite{rombach2022high} is formulated as follows:
\begin{equation}
   \mathcal{L}_{\text{LDM}} := \mathbb{E}_{\varepsilon(x), \varepsilon \sim \mathcal{N}(0,1), t} \left[ \left\| \varepsilon - \varepsilon_{\theta}(z_t, t) \right\|^2_2 \right],
\label{equ:ldm}
\end{equation}
where $\varepsilon_{\theta}(z_t, t)$ denotes a denoising autoencoders (time-conditional UNet) and $\theta$ contains the corresponding weights, where $t = 1...T$. 

\subsection{Knee Joint Identity Prior and Injection}

To persist a subject's identity information in generating the corresponding 12-month X-ray scan, an identity prior module is devised to extract the identity-sensitive patterns. 
Specifically, we denote the identity representation for the $i^{th}$ subject joint as $\mathbf{F}_{\text{ID}}^i \in \mathbb{R}^d$, where $d$ indicates the dimension of the identity representation. A CNN-based network $f(\cdot)$ is designed to obtain the identity representation. Mathematically, we have:
\begin{equation}
\mathbf{F}_{\text{ID}}^i = f(\mathbf{I}_{\text{v00m}}^i). 
\label{equ:identity_prior}
\end{equation}
The identity information guides and regularizes the diffusion process to not only focus on the target visual domain distribution but also maintain necessary information from the subjects. 

For the supervision of $f$ to formulate identity information, it is proposed to conduct training in a contrastive learning manner. A triplet loss is introduced for this purpose as follows:
\begin{equation}
\mathcal{L}_{\text{Triplet}} = \sum \left[ \left\| f(\mathbf{I}_{\text{v00m}}^i) - f(\mathbf{I}_{\text{v12m}}^i) \right\|_2^2 - \left\| f(\mathbf{I}_{\text{v00m}}^i) - f(\mathbf{I}_{\text{v00m}}^k) \right\|_2^2 + \alpha \right]_+,
\label{equ:triplet_loss}
\end{equation}
where a triplet is formulated as $(\mathbf{I}_{\text{v00m}}^i, \mathbf{I}_{\text{v12m}}^i, \mathbf{I}_{\text{v00m}}^k)$: $\mathbf{I}_{\text{v00m}}^i$ indicates the anchor of current subject's joint; $\mathbf{I}_{\text{v12m}}^i$ refers to a positive sample from the same subject's joint; $\mathbf{I}_{\text{v00m}}^k$ is a negative sample with $i\neq k$; and $\alpha$ is a margin enforcing between positive and negative pairs. 
Note that the number of triplets $M$ used in the training process is defined in the experiment implementations. 

The identity information acts as a guide and regulator for the diffusion process, ensuring that it not only aligns with the target visual domain distribution but also preserves essential information pertaining to the subjects. To inject the identity into the diffusion process, 
%The identity representations $\mathbf{F}_{\text{ID}}$ obtained are injected into the diffusion-based model to persist the patient's identity. In particular, 
a cross-attention mechanism is applied between the features from the UNet layers in $\epsilon_\theta$ and $\mathbf{F}_{\text{ID}}$. For the $t^\text{th}$ diffusion step and $j^\text{th}$ layer, the attention can be formulated as follows: 
\begin{equation}
\mathbf{A}_j = \text{Softmax}(\dfrac{{\mathbf{Q}}{\mathbf{K}^{\intercal}}}{\sqrt{d}})\mathbf{V},
% \begin{equation}
\mathbf{Q} = \phi_j(z_t)\mathbf{M}^Q_j \text{, } \mathbf{K} = \mathbf{F}_{\text{ID}}^{i}\mathbf{M}^K_j \text{, and } \mathbf{V} = \mathbf{F}_{\text{ID}}^{i}\mathbf{M}^V_j \text{, }
\label{equ:cross_att_qkv}
\end{equation}
% \begin{equation}
% \text{Cross Attention} = \text{Softmax}(\dfrac{{\mathbf{Q}}{\mathbf{K}^{\intercal}}}{\sqrt{d}})\mathbf{V}.
% \label{equ:cross_att}
% \end{equation}
where $\mathbf{M}^{\circ_j}$ represents a projection matrix containing trainable weights with a projection dimension $d$, and $\phi_j(z_t)$ indicates the (flattened) intermediate representation of the UNet with $j^\text{th}$ layer. The UNet can be formulated by: 
\begin{equation}
\phi_{j+1}(z_t) = \phi_j(z_t) + \mathbf{A}_j,
\label{equ:unet}
\end{equation}

To this end, the IC-RDN can be learnt via the following scheme:
\begin{equation}
   \mathcal{L}_{\text{LDM-Identity}} := \mathbb{E}_{\varepsilon(x), \varepsilon \sim \mathcal{N}(0,1), t} \left[ \left\| \varepsilon - \varepsilon_{\theta}(z_t, t, f(\mathbf{I}_{\text{v00m}}^i)) \right\|^2_2 \right].
\label{equ:Knee_ICGN}
\end{equation}

\subsection{Generation-Guided KOA Progression Prediction} \label{subsec:gen_guid_predict}

In addition to generating 12-month X-ray image ${\hat{\mathbf{I}}_{\text{v12m}}^i}$ for prognosis, a singular progression severity grade is predicted as well based on the baseline visit X-ray image ${{\mathbf{I}}_{\text{v00m}}^i}$ and the generated 12-month visit X-ray image ${\hat{\mathbf{I}}_{\text{v12m}}^i}$. 
Specifically, a two-stream CNN backbone with two branches $h_{\text{v00m}}(\cdot)$ and $h_{\text{v12m}}(\cdot)$ is applied for the two images to produce the associated representations:
\begin{equation}
   \mathbf{P}_\text{v00m}^i=h_{\text{v00m}}({{{\mathbf{I}}_{\text{v00m}}^i}}),  \mathbf{P}_\text{v12m}^i=h_{\text{v12m}}({\hat{\mathbf{I}}_{\text{v12m}}^i}). 
\end{equation}
The two representations are then concatenated, denoted by $\Vert$, and used in a multi-layer perceptron (MLP) classifier $g(\cdot)$ to produce the predicted progression severity grade ($\hat{y}_{KL}^i$), which can be formulated as:
\begin{equation}
   \hat{y}_{KL}^i = g( \mathbf{P}_\text{v00m}^i \Vert \mathbf{P}_\text{v12m}^i)
\label{equ:downstream_cls}
\end{equation}

A cross-entropy loss is leveraged in the training process to guide the prediction training process, which can be formulated as: 

\begin{equation}
\mathcal{L}_{\text{CE}} = -\sum_{i=1}^{N} y_{KL}^i \log \hat{y}_{KL}^i
\label{equ:ce_loss}
\end{equation}
where $N$ indicates the total number of joints in the dataset. 

\section{Experimental Results and Discussion} \label{sec:gen_results}

\subsection{Dataset and Pre-Processing}

The longitudinal X-ray images used in this study were sourced from the Osteoarthritis Initiative (OAI) \cite{lester2008clinical}, which is a multi-centre and longitude study containing chronological medical imaging scans and clinical assessment records for the subjects up to 96 months. This public dataset enables researchers to explore the natural progression of osteoarthritis, identify risk factors, and evaluate biomarkers for grading the KOA severity and progression. In this study, the patients with both baseline and 12-month visits and available 12-month severity KL grades are selected. Due to some participants opting out of the follow-up study and missing recording of clinical severity grading outcome, not all have recordings for the 12-month visit image and severity KL grades. Consequently, 7,923 joints have both the KL grade available and radiographic recordings from both the baseline and 12-month visits. As the demographic information shown in Table \ref{tab:demographic}, the amount of joints with changed severity grades among 12 months is much smaller than that of without changes, where only 6.15\% joints show a changed severity grade. However, across these joints without the severity grade changes, the visual structural details do not remain the same, which also shows the importance of the proposed multifaceted progression predicting method. The dataset was randomly divided among subjects with a distribution of 70\%, 10\%, and 20\% for training, validation, and testing, respectively. 

\begin{table}[htbp]
  \centering
  \caption{Statistics of patients, joints and KL grades in OAI longitudinal visits}
    \begin{tabular}{c|ccc}
    \hline
    Longitudinal Visits & Patients & Available Joints & Joints with KL Grades \\
    \hline 
    Baseline Visit & 4,551   & 9,102   & 8,953 \\
     12-Month Visit  & 4,786   & 9,418   & 8,251  \\
     \hline
    Longitudinal KL Changed &   \multicolumn{3}{c}{487 (6.15\%)} \\
    Longitudinal KL Unchanged & \multicolumn{3}{c}{7,436 (93.85\%)} \\
     \hline
    Both Visits &   \multicolumn{3}{c}{7,923}   \\
    \hline
    \end{tabular}%
  \label{tab:demographic}%
\end{table}%

As the raw data from the OAI contains bilateral knee joints and includes almost all knee bones around the joint area, knee joint localisation methods are leveraged to extract the knee joint area individually \cite{lindner32consortium}. After detecting the knee joint edge, a 224$\times$224 region surrounding the middle of the joint was extracted from the raw data for modelling. The remaining transformation approaches for the joint images are followed \cite{chen2019fully}, including adjusting the brightness, contrast and saturation of the images.

\subsection{Implementation Details}
\label{subsec:implement}

A pretrained ResNet 18 on ImageNet~\cite{deng2009imagenet} is adopted as the backbone classification model for identity detection, which extracts the identity features from the baseline X-ray scans. Within Identity Prior's contrastive learning framework, we have designated the number of triplets $M$ as 7,000 for training and 1,000 for testing. 
The Identity Prior training phase is fine-tuned using the Adam optimizer with a learning rate of 0.01, and the batch size for this phase is established at 128. 
For the training process of the diffusion-based generative module,  we set the number of steps to 1,000,000 across 400 epochs. 
The Adam optimizer is employed here as well, with a learning rate set at $1\times 10^{-4}$. The batch sizes for training and testing are 64 and 8, respectively. 

For the backbone models in the generation-guided KOA progression prediction module, various CNN methods are adopted, such as VGG 16, VGG 19, ResNet 18, ResNet 50, and DenseNet 121, for a comprehensive comparison \cite{simonyan2014very, he2016deep, huang2017densely}. Specifically, the backbone prediction models for both branches, corresponding to $\mathbf{I}_{\text{v00m}}$ and $\hat{\mathbf{I}}_{\text{v12m}}$, share the same architecture. In addition, the output features dimension from these backbone models is set to 128. Then the MLP classifier $g(\cdot)$ takes the input dimension of 256, which contains two linear layers, an activation layer, and a dropout layer. 
The training process is conducted with a batch size of 16. An SGD optimizer is utilized with the learning rate of $5\times 10^{-3}$ and a weight decaying factor of $5\times 10^{-3}$ for every 5 epochs.
The entire training regimen is executed using PyTorch 2.0.1 on an NVIDIA RTX A6000 with 48 GB of video memory.

\begin{table*}[!htbp]
  \centering
  \caption{Comparison of KOA progression grading performance for different visual generation methods.}
    \begin{tabular}{c|c|ccccc}
    \hline
    \multirow{2}{*}{Method} & \multirow{2}{*}{IS} & \multicolumn{5}{c}{Progression Grading Models - Accuracy (\%)} \\
    \cline{3-7}
          &  & VGG-16 & VGG-19 & ResNet-18 & ResNet-50 & DenseNet-121 \\
                   \hline 
     Baseline Image Only & - & 66.3 & 63.3 & 60.8 & 60.9 & 62.5  \\
     Cycle GAN \cite{zhu2017unpaired} & 1.28 & 62.5 & 64.3 & 60.3 & 61.2 & 63.6 \\
     I2I Diffusion Model & \textbf{2.02} & 62.4 & 62.9 & 60.4 & 61.3 & 62.7 \\
     CXR-IRGen \cite{irgen2024} & 1.76 & 61.1 & 62.7 & 60.3 & 59.1 & 61.6 \\
     PITET GAN \cite{petit2024} & 1.61 & 63.3 & 63.7 & 59.6 & 60.7 & 62.2 \\
     \hline
     IC-RDN (ours) & 1.98 & \textbf{66.8} & \textbf{65.1} & \textbf{61.1} & \textbf{61.7} & \textbf{64.4} \\
    \hline 
    \end{tabular}%
  \label{tab:overall_performance}%
\end{table*}%

\subsection{Quantitative Performance on Visual Prognosis}

The evaluation of the generated future X-ray images encompasses both quantitative and qualitative assessments. Quantitatively, we engage in analysis through the conventional generation task metrics, inception score and a downstream task: the prediction of KOA progression severity grade. For using the inception score in this study, we use the Inception-v3 network trained on the OAI 12-month visit X-ray dataset to replace the weights pre-trained on the ImageNet~\cite{deng2009imagenet} as the knee X-ray images are different from the natural images at both category and structural level. As shown in Table \ref{tab:overall_performance}, the X-ray images generated by our proposed IC-RDN illustrate a similar IS with the X-ray images generated by the image-to-image diffusion model and outperform the X-ray images generated by Cycle-GAN \cite{zhu2017unpaired} and other SOTA generative models \cite{irgen2024, petit2024} significantly. As for the similar IS between the proposed IC-RDN and the image-to-image diffusion model, the results show similar performance on image quality and diversity. However, 
the images generated by the I2I diffusion model show a minor contribution to predicting the KOA progression severity grades, and the challenge of identity inconsistency remains in the generated X-ray images from the I2I diffusion model. The detailed comparison for identity consistency is shown on Sec. \ref{subsec:identity_results}. 

Although the inception score is able to show the quality and diversity of the generated future images, the downstream task of predicting the KOA progression severity grade can be more relevant to the value of how the generated future X-ray images contribute to exploring the KOA progression patterns. We have introduced the approach to predict KOA severity progression (Sec. \ref{subsec:gen_guid_predict}), leveraging both the generated future X-ray image and the baseline X-ray image to determine a progression severity grade. 
% As directly using the generated images in real clinical practice can be dangerous due to computer hallucinations, we designed this downstream task to support the specialist's decisions. 
In addition, as stated in Sec. \ref{subsec:implement}, various CNN-based backbones are deployed in the generation-guided KOA progression prediction module. The comprehensive results among these backbone models are shown in Table. \ref{tab:overall_performance}. 

Specifically, the predicted 12-month progression X-ray images generated by Cycle GAN, the image-to-image diffusion model, the recent SOTA generative methods, and the proposed IC-RDN are used in the generation-guided KOA progression grade prediction module separately. Accuracy is used as the evaluation method to measure the performance of predicting the progression grades. Among all backbone models used in this module, predicting the progression severity grades using X-ray images from baseline visit and generated by the proposed IC-RDN shows significantly outperformed results. In detail, the image-to-image diffusion model shows a limited contribution to providing the multifaceted KOA progression patterns. And the Cycle GAN illustrates the minor ability to enhance the performance of predicting the KOA progression severity grades. Therefore, the quantity evaluation demonstrates the added value of incorporating generated future X-ray images in enhancing the accuracy of KOA progression severity grade predictions.

\begin{figure*}[!htb]
\centering
\includegraphics[width=1\textwidth]{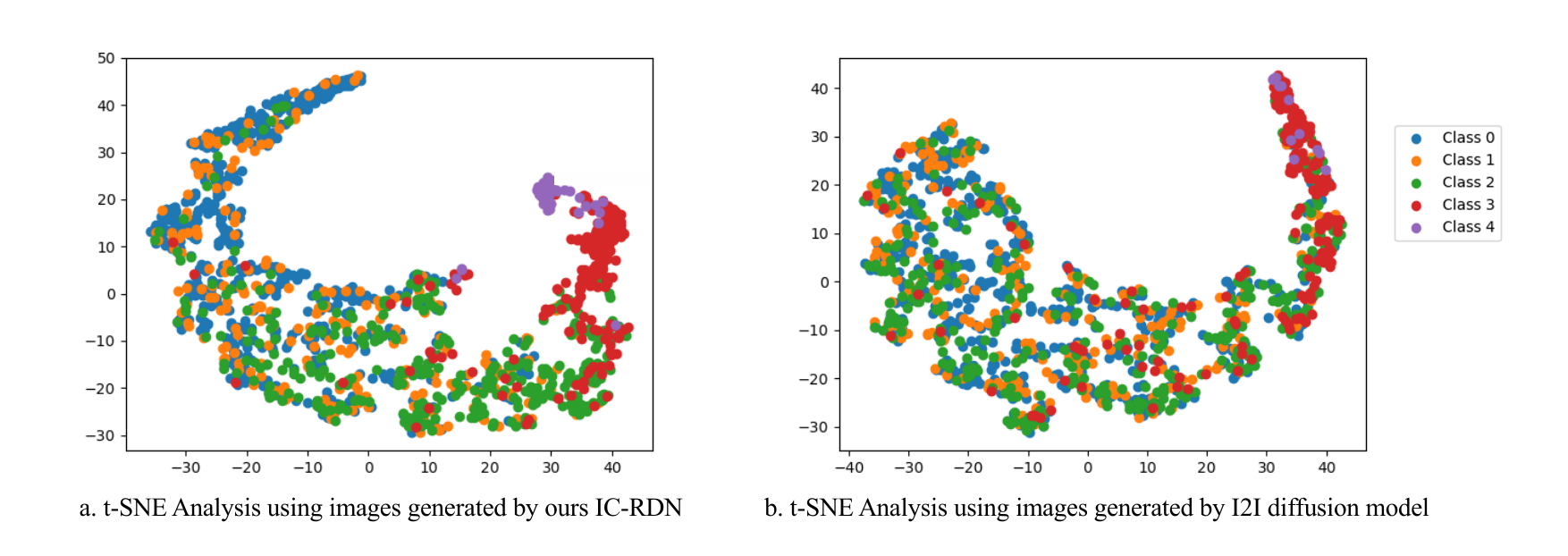}
\caption{Illustration of t-SNE analysis between IC-RDN (ours) and I2I diffusion model. }
\label{fig:tsne_analy}
\end{figure*}

To illustrate the performance of generated X-ray images contributing to predicting the KOA progression grades, the t-Distributed Stochastic Neighbor Embedding (t-SNE) technique, a sophisticated machine learning algorithm designed for the task of high-dimensional data visualization, is utilised in this study. In detail, to reduce the complex, multi-dimensional data into a two-dimensional space, t-SNE converts the high-dimensional distances between data points into conditional probabilities that represent similarities. Specifically, given a predicted X-ray image generated by the proposed IC-RDN, the trained VGG 19 network is used to produce a feature map corresponding to the KOA progression grade patterns. Then, these representations are visualised using t-SNE to show the related grades spatially. As shown in Fig. \ref{fig:tsne_analy}, the nodes with different colours indicate the corresponding grades. In Fig. \ref{fig:tsne_analy}.a, clear clusters for the KL grades of 0, 3, and 4 are located on the left and right sides. The green and yellow nodes referred to KL grades 1 and 2, show that the module for predicting the KOA progression severity grade has a minor ability to categorise these two classes, which is the challenge among different KOA progression prediction and severity grading methods \cite{chen2019fully, wu2023self}. However, in Fig. \ref{fig:tsne_analy}.b, the nodes for different grades are mixed and the cluster centres are hard to identify. 

\begin{figure*}[!htb]
\centering
\includegraphics[width=1\textwidth]{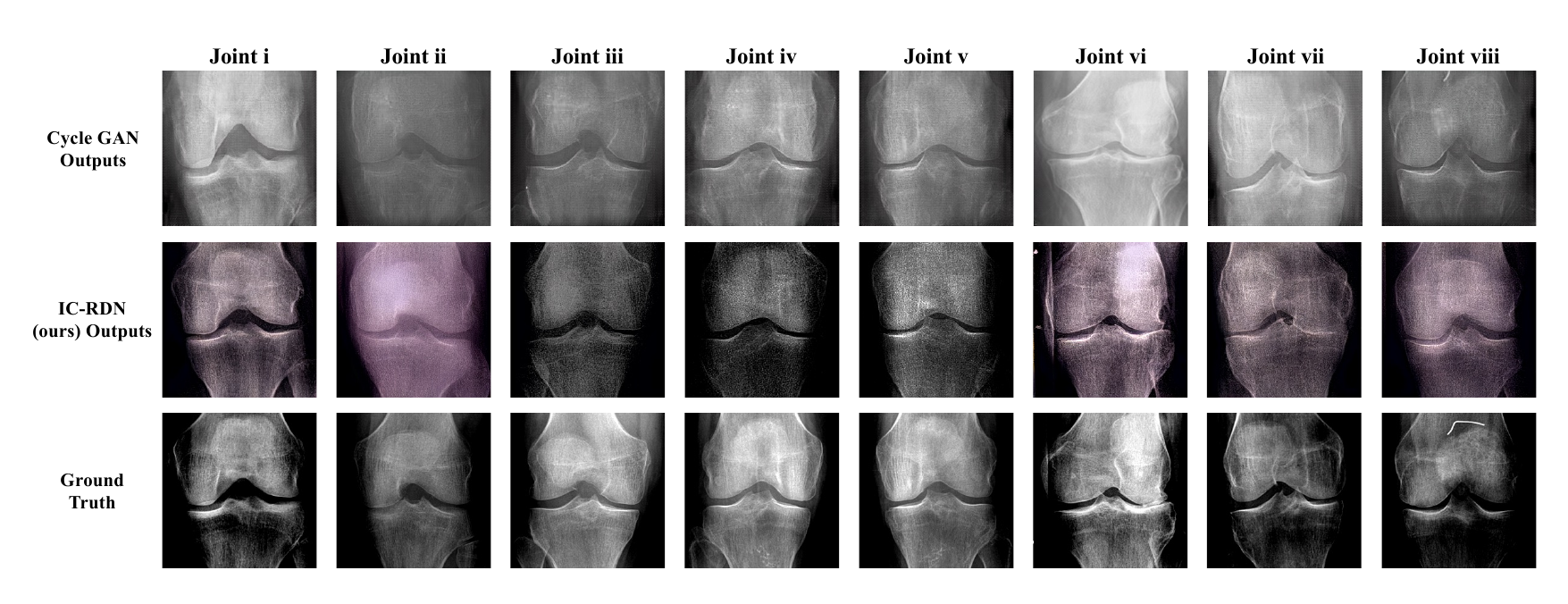}
\caption{Comparison of generated X-ray images from Cycle GAN \cite{zhu2017unpaired} and our proposed IC-RDN on knee joints i-viii. }
% (corresponding to  joint ID 9926321\_2, 9871731\_2, 9871731\_1, 9897629\_1, 9897629\_2, 9048789\_2, 9026934\_1, 9402978\_2)
\label{fig:GANvsDiffusion}
\end{figure*}

\begin{figure*}[!htb]
\centering
\includegraphics[width=1\textwidth]{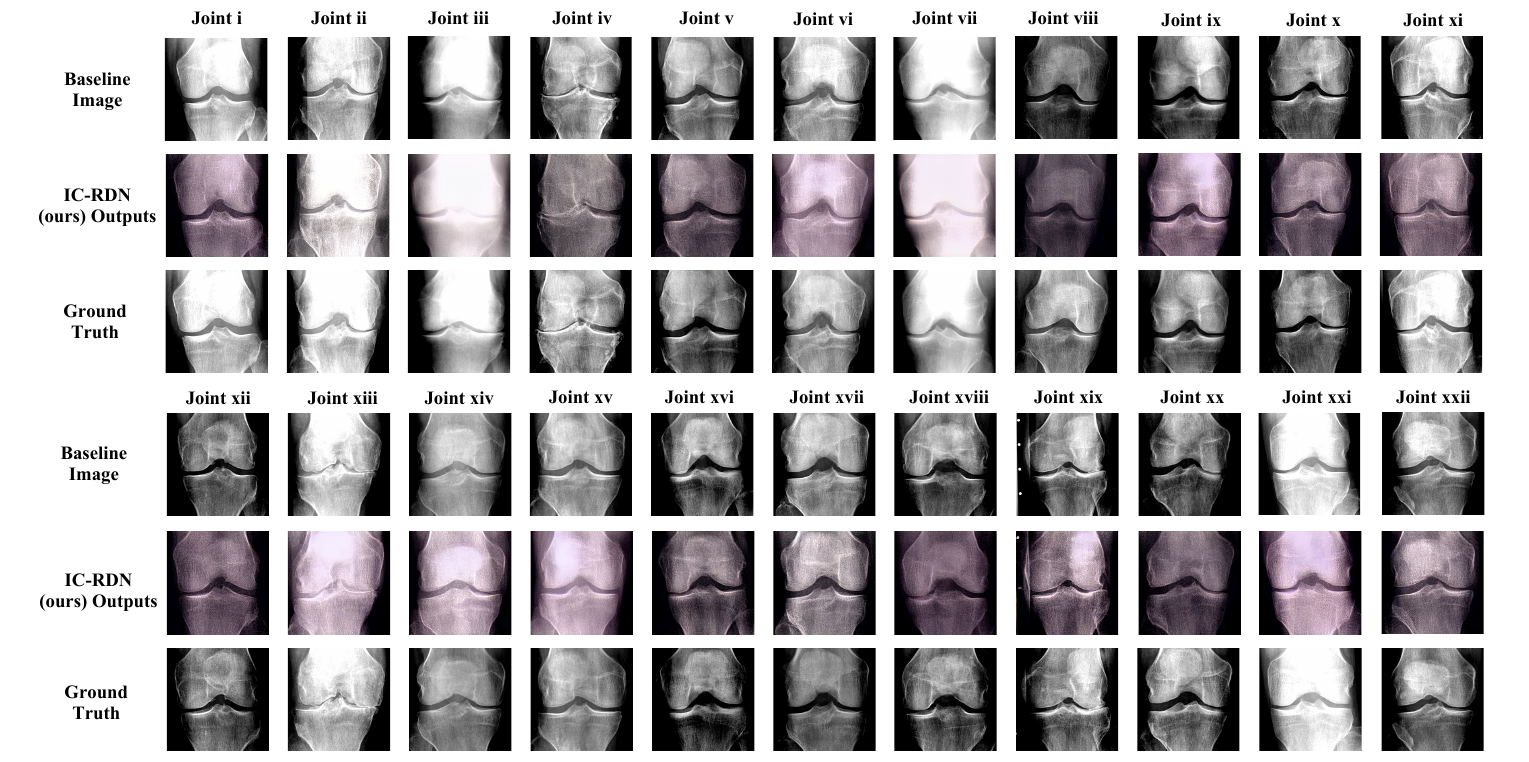}
\caption{Generative results from IC-RDN among different patients' joints from i to xxii. The detailed joint structure comparison between baseline visit images, ground truth and results from IC-RDN shows our model's ability to generate progression patterns. }
\label{fig:gen_results}
\end{figure*}

\subsection{Qualitative Performance on Visual Prognosis}
\label{subsec:qualitatuve_perf}

In terms of qualitative analysis, the comparison between X-ray images synthesized by the Cycle GAN and those produced by the proposed IC-RDN offers a compelling visual assessment, as depicted in Fig. \ref{fig:GANvsDiffusion}. The comparative study highlights notable distortions in bone structures emanating from the Cycle GAN approach, particularly in the articulation of joint spaces. In stark contrast, our proposed IC-RDN exhibits a pronounced fidelity in preserving the integrity of key bone edges and the intricacy of inner textures, thereby underscoring the superior capability of IC-RDN in maintaining structural information.

Further elucidation is provided through an expanded corpus of X-ray images generated by the IC-RDN, which is shown in Fig. \ref{fig:gen_results}. This compilation serves not only to reinforce the initial observations regarding the preservation of critical structural details but also to demonstrate the consistent performance of the IC-RDN across a diverse set of examples. The consistent replication of complex anatomical features by the IC-RDN model is indicative of its sophisticated understanding of underlying bone structures, which is paramount for illustrating the multifaceted KOA progression patterns.

Moreover, this comparison elucidates the inherent limitations of traditional GAN-based approaches in contexts where the preservation of exact structural features is paramount. The propensity for distortion within Cycle GAN-generated images can significantly impact the clinical utility of such synthesized X-ray images, particularly in monitoring the KOA progression. In contrast, the proposed IC-RDN shows its adeptness at capturing and reproducing the nuanced details of bone structure suggesting its potential as a multifaceted KOA progression prediction outcome.

\begin{figure*}[!htb]
\centering
\includegraphics[width=1\textwidth]{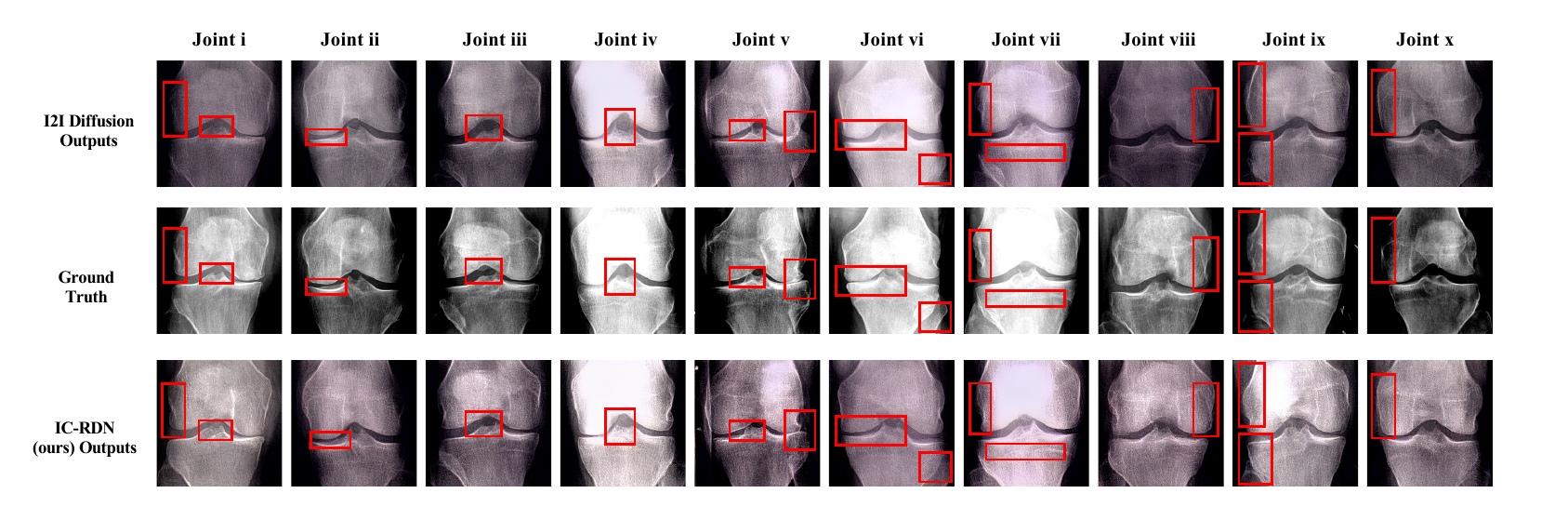}
\caption{Examples of generated X-ray images with identity consistency for joint i-x. The first line (outputs from the Image-to-Image Diffusion Model) shows the generative results without Identity Prior. The second line indicates the ground truth images. The third line refers to the results generated by our IC-RDN approach.}
% (corresponding to the joint ID of 9002817\_1, 9002817\_2, 9009927\_1, 9036770\_2, 9048789\_2, 9096724\_2, 9129270\_1, 9142551\_1, 9152569\_1, 9319366\_2 from OAI)
\label{fig:identity_results}
\end{figure*}

\subsection{Analysis on Identity Consistency}
\label{subsec:identity_results}

To further evaluate the efficacy of the identity prior module, we delve into a comprehensive comparison involving the X-ray images generated by the I2I diffusion model and the proposed IC-RDN. The visual examples presented in Fig. \ref{fig:identity_results} meticulously outline the distinctions between these methods alongside the actual 12-month visit X-ray images, where the key differences are highlighted by the bounding boxes. Initially, we observe the generated X-ray images devoid of identity persistency, which starkly contrasts with the subsequent imagery. 
% Following this, the 12-month visit X-ray images are showcased, succeeded by the X-ray images synthesized through our innovative IC-RDN approach. 
A noteworthy observation is the preservation of identity features in the X-rays produced by our IC-RDN, as evidenced by the distinct texture and precision of bone edge structures. This preservation extends to intricate details within the knee joint, such as the orientation of the menisci and the nuanced protrusions along the bone edges, which are critical for maintaining identity persistency.

\subsection{Ablation Study}

To better evaluate the contribution of the proposed Identity Prior module, an ablation study is conducted by removing the Identity Prior and only the Diffusion backbone remaining. The results in Table \ref{tab:ablation} demonstrate the effectiveness of the Identity Prior module with a superior longitude prediction accuracy among all selected classification methods. 

\begin{table*}[!htbp]
  \centering
  % \small
  \caption{Ablation study of the proposed IC-RDN}
    % \resizebox{0.47\textwidth}{!}{
    \begin{tabular}{c|c|ccccc}
    \hline
    \multirow{2}{*}{Method} & \multirow{2}{*}{IS} & \multicolumn{5}{c}{Progression Grading Models - Accuracy (\%)} \\
    \cline{3-7}
          &  & VGG16 & VGG19 & ResNet18 & ResNet50 & DenseNet121 \\
                   \hline 
     IC-RDN w/o Identity Prior & 1.72 & 61.6 & 61.7 & 60.2 & 61.0 &  62.5 \\
     IC-RDN (ours) & \textbf{1.98} & \textbf{66.8} & \textbf{65.1} & \textbf{61.1} & \textbf{61.7} & \textbf{64.4} \\
    \hline 
    \end{tabular}%
    % }
  \label{tab:ablation}%
\end{table*}

\subsection{Limitations}

The major limitation of this study regards the generation quality. Directly using X-ray scans generated by IC-RDN for a KOA severity grading showcases a suboptimal result compared with using the baseline visit X-ray to predict the KOA progression grade directly. Though the predicted 12-month scan can enhance the KOA severity prediction with using the baseline visit jointly. In addition, computer hallucinations \cite{gunjal2024detecting} are a crucial issue among large model fields. Particularly, the analysis regarding the disease-specific changes is insufficient. Therefore, careful consideration of applying the generative method to actual clinical usage is needed. 

\section{Conclusion} \label{sec:conclusion}

In this study, a generative-based method, IC-RDN is proposed for KOA prognosis, which forecasts future X-ray scans for joints and singular progression KOA severity grades. This obtains multifaceted KOA progression patterns for understanding and explaining KOA prognosis.
Specifically, an identity prior module for the diffusion and a downstream generation-guided progression prediction module are introduced for regularising and guiding the diffusion to focus more on the clinical nuances related to the prognosis and producing a more precise KOA severity progression grading. 
% Compared to a conventional image-to-image generative model, identity priors regularize and guide the diffusion to focus more on the clinical nuances related to the prognosis, based on a contrastive learning strategy. The progression prediction module utilizes both forecasted and baseline knee scans for a more precise KOA severity progression grading.  
Extensive experiments on a widely used public dataset, OAI, demonstrate the effectiveness of the proposed method.
% In our future study, the features related to identity and KOA severity will be disentangled for more specified OA pattern exploration. 
In our future studies, the disentanglement of the representations related to the identity and KOA severity will be explored for fine-grained guidance in the X-ray image generation process. Meanwhile, the multimodal patterns among different medical imaging scans (\eg X-ray, MRI, and CT) and patients' demographic information will be involved in producing a fine-grained KOA progression assessment outcome.

\section*{Acknowledgements}
This study was partially supported by Australian Research Council (ARC) grant DP210102674. The OAI \cite{lester2008clinical} is a public-private partnership comprised of ﬁve contracts (N01-AR-2-2258; N01-AR2-2259; N01-AR-2- 2260; N01-AR-2-2261; N01-AR-2-2262) funded by the National Institutes of Health, a branch of the Department of Health and Human Services, and conducted by the OAI Study Investigators.

\bibliographystyle{splncs04}
\bibliography{main}
\end{document}